\documentclass[conference]{IEEEtran}
\usepackage{times,amsmath,epsfig,amssymb,graphicx,amsfonts}
\usepackage{subfigure}
\usepackage{psfrag}
\usepackage{ifpdf}

\title{A General Analog Network Coding for Wireless Systems with Fading and Noisy Channels}

\author{\IEEEauthorblockN{M. Amin Rahimian\IEEEauthorrefmark{1}\IEEEauthorrefmark{2}, Ali Ayremlou\IEEEauthorrefmark{1}\IEEEauthorrefmark{3}, Farokh Marvasti\IEEEauthorrefmark{1}\IEEEauthorrefmark{4}}
\IEEEauthorblockA{\IEEEauthorrefmark{1}Advanced Communications Research Institute, Sharif University of Technology 
Tehran, Iran\\
\IEEEauthorrefmark{2}Email: rahimian.amin@gmail.com, \IEEEauthorrefmark{3}Email: a\_ayremlou@ee.sharif.edu, \IEEEauthorrefmark{4}Email: marvasti@sharif.edu}
}

\begin{document}
\maketitle

\begin{abstract}
It has been recently brought into spotlight that through the exploitation of network coding concepts at physical-layer, the interference property of the wireless media can be proven to be a blessing in disguise. Nonetheless, most of the previous studies on this subject have either held unrealistic assumptions about the network properties, thus making them basically theoretical, or have otherwise been limited to fairly simple network topologies. We, on the other hand, believe to have devised a novel scheme, called Real Amplitude Scaling (RAS), that relaxes the aforementioned restrictions, and works with a wider range of network topologies and in circumstances that are closer to practice, for instance in lack of symbol-level synchronization and in the presence of noise, channel distortion and severe interference from other sources. The simulation results confirmed the superior performance of the proposed method in low SNRs, as well as the high SNR limits, where the effect of quantization error in the digital techniques becomes comparable to the channel. 
\end{abstract}


\section{Introduction}\label{introduction}

\begin{figure*}[ht]
\centering
\subfigure[Routing]
{\label{fig:routing}\includegraphics[width=59mm]{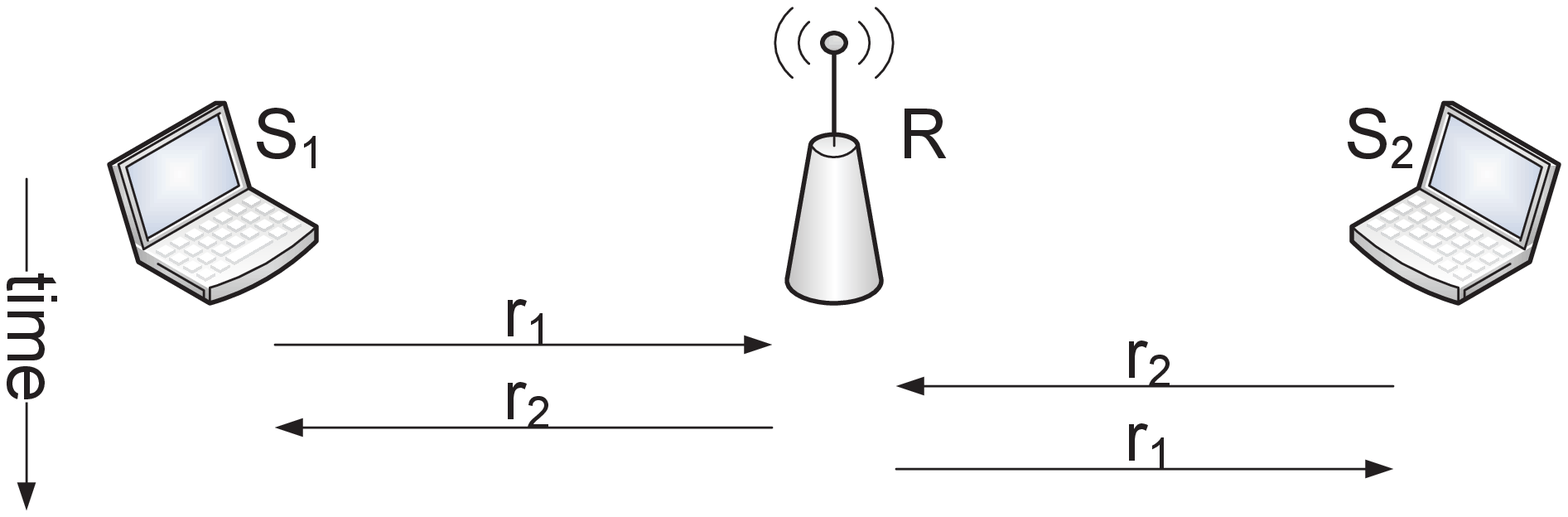}}
\subfigure[Algebraic Coding]
{\label{fig:algebraicoding}\includegraphics[width=59mm]{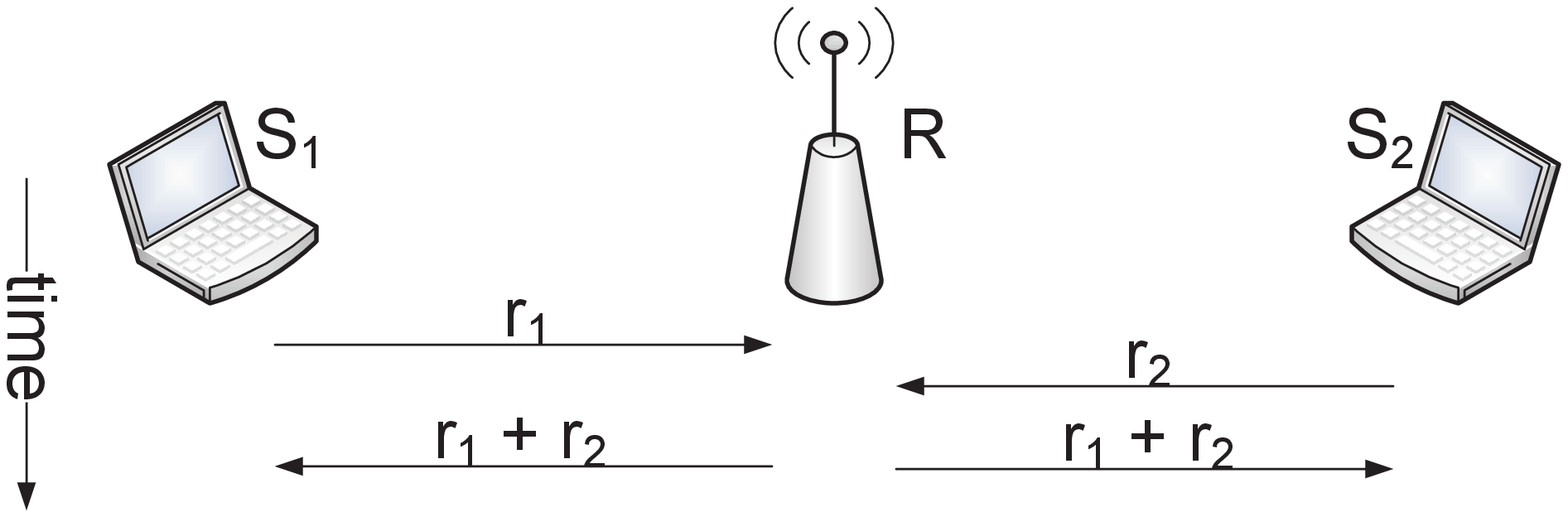}}
\subfigure[Arithmetic Coding]
{\label{fig:Arithmeticoding}\includegraphics[width=59mm]{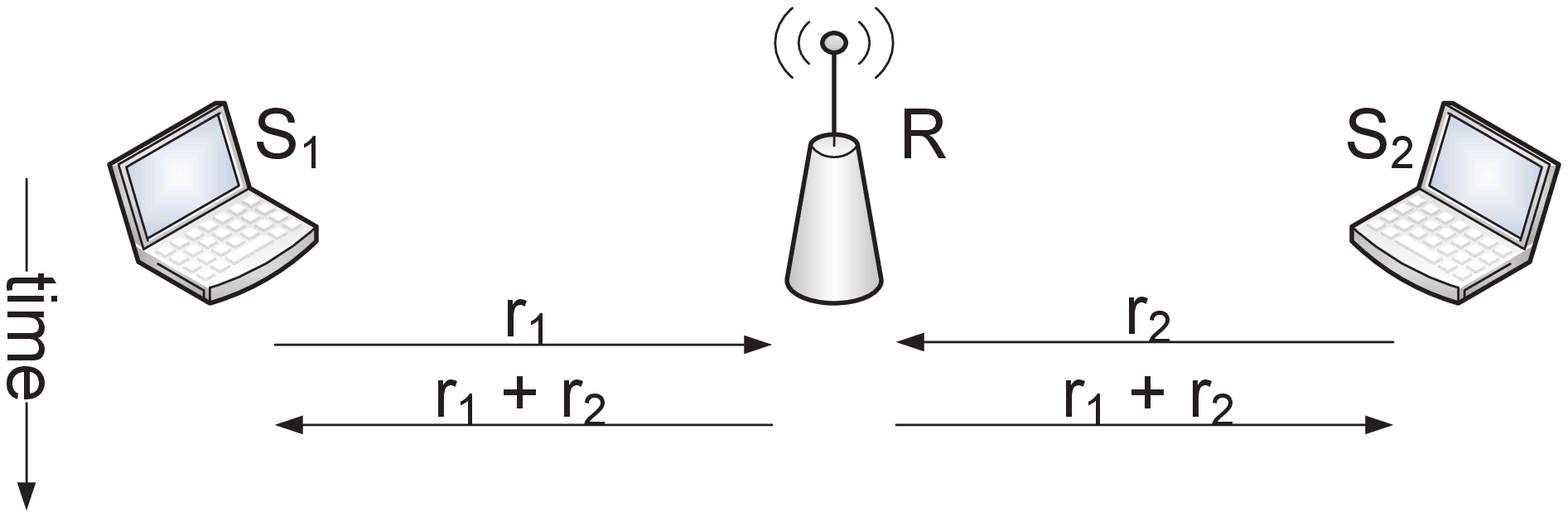}}
\caption{Nodes $S_1$ and $S_2$ wish to exchange over a wireless network their messages, respectively $r_1$ and $r_2$, through node $R$ as relay. (a) shows the ``usual" routing scheme, taking a total of 4 time-slots. (b) notes that R can instead broadcast the sum $r_1+ r_2$, resulting in a total of 3 time-slots. (c) implements an Arithmetic Network Code, where $S_1$ and $S_2$ transmit messages to R simultaneously, then R simply broadcasts $r_1+r_2$. This requires just 2 time-slots.}
\vspace{-.1in}
\label{fig:figure1}
\end{figure*}

Consider the three-node chain network in Fig. \ref{fig:figure1}. Nodes $S_1$ and $S_2$ are the source nodes which are going to communicate with each other. In particular, we consider two real numbers $r_1$ and $r_2$, which $S_1$ and $S_2$ are going to exchange, respectively. Since the source nodes are out of the radio range of each other, they cannot communicate without a relay node. This simple three-node network forms the canonical example for wireless network coding and has been subject to extensive previous studies \cite{shish}. In the traditional IEEE 802.11 scheme, $S_1$ would send its packet to the router $R$, which then forwards it to $S_2$. Similarly, $S_2$ sends its packet to $R$, which again forwards it to $S_1$. This way, in order to exchange two packets, the traditional approach takes up $4$ time-slots.

Through the exploitation of network coding the same goal can be achieved in fewer time slots. In particular, $S_1$ and $S_2$ can send their packets to the router, one after the other in two consecutive time slots; the router then $XOR$s the two packets and broadcasts the $XOR$-ed version. After both $S_1$ and $S_2$ received the $XOR$-ed version in the third time-slot, $S_1$ recovers $S_2$'s packet by removing its own message from the combination, i.e. by $XOR$-ing again with its own message, which $S_1$ knows by the virtue of having sent it and vice versa. Thus, network coding reduces the number of time slots from 4 to 3. The freed slot can be used to send new data, improving wireless throughput \cite{noh,sizdah}.
However, it is still possible to decrease the number of time-slots. $S_1$ and $S_2$ could transmit their packets simultaneously, allowing their transmissions to interfere at the router. This consumes a single time slot. Due to interference, the router receives the arithmetic sum of $S_1$'s and $S_2$'s signals, $s_1(t) + s_2(t)$, and thus the router cannot decode the received signal. Nonetheless, the router can simply amplify and forward the received interfered signal at the physical layer without any attempt to decode it. This consumes a second time slot. Since sources know the packet they transmitted by the virtue of having sent it, they can substract it from the recieved interfered signal. Such a scheme of transmission is dubbed physical-layer network coding (PNC). According to what was mentioned, PNC enables us to reduce the required number of time slots from 4 to 2, thus doubling the wireless throughput \cite{yek,do}.

Our work builds particularly upon the previous literature \cite{yek,do}; however, the approach is fundamentally different from both of them. In particular, we consider a noisy channel with fading and distortion, and make no assumption about the network synchronization,which makes our work particularly significant, from a practical point of view. Furthermore, our method boasts a great potential for extensions to larger networks with random topologies.

we have considered two methods for the implementation of the physical-layer network coding idea: the digital encoding and Real Amplitude Scaling (RAS). For both methods -- digital encoding and RAS -- the transmitted data packets are comprised of two separate sections. The first part, called signature, is a PN-sequence. The term PN-sequence refers to a random sequence of equiprobable binary symbols, called chips. A distinct PN-sequence is assigned to each of the source nodes and all of the PN-sequences have the same length (number of chips). Moreover, it is assumed that every PN-sequence is known to all source nodes across the network. The second section in each data pack differs based on the approach adopted. As for the digital encoding approach, the second section is a digitally encoded version of the real number that is to be transmitted. In the RAS method, on the other hand, the second part is also a PN-sequence. Its amplitude, however, is scaled according to transmitted data.

The simulated wireless channel is an additive, white, Gaussian noise (AWGN) channel which introduces a uniformly distributed random delay, as well as a random attenuation, observing Rayleigh distribution, for each block of data. Accordingly a message $r_1(t)$ transmitted by $S_1$, in the context of the topology depicted in Fig. \ref{fig:Arithmeticoding} will have undergone a net random delay ${\tau}_{s_1r} + {\tau}_{rs_2}$ and a total random attenuation of $g_{s_1s_2} = h_{s_1r}.h_{rs_2}$, by the time it is received at $S_2$. Same discussion applies to $S_2$'s own message, $r_2$, which will interfere with $S_1$'s message, while being received by the Relay node slot and will then be forwarded together with $S_1$'s message back to $S_2$ by the Relay node. The message received by $S_2$ at the end of the second time-slot can be formulated as:

\begin{eqnarray}
y_{2}(t) &=& (h_{s_1r}.h_{rs_2})r_1(t-({\tau}_{s_1r} + {\tau}_{rs_2})) + \nonumber 
\\&& (h_{s_2r}.h_{rs_2})r_2(t-({\tau}_{s_2r} + {\tau}_{rs_2}))\\
&=& {g}_{s_1s_2}.r_{1}(t-({\tau}_{s_1r} + {\tau}_{rs_2})) + \nonumber 
\\&& g_{s_2s_2}.r_{2}(t-({\tau}_{s_2r} + {\tau}_{rs_2}))
\label{channel}
\end{eqnarray}

According to what was mentioned, $g_{s_is_j}$ denotes the overall change that the amplitude of a message transmitted from source node $S_i$ will undergo before being received at the source node $S_j$.

The rest of this paper is organized as follows: Section \ref{proposed}, then presents the digital encoding and Real Amplitude Scaling (RAS) which constitutes the major focus of this paper. The operations of the proposed coding schemes were simulated, and the results are presented in section \ref{sec:sim}. Section \ref{sec:conc} concludes this paper. 
\section{Proposed Wireless Network Coding Schemes}\label{proposed}

\subsection{Digital Encoding}\label{sec:digitEnc}

As already highlighted in the previous section, the data packets in this method are comprised of two separate parts: PN-sequence and a digital sequence of data bits that contain the actual message that is to be transmitted. In the receivers' side, each of the source nodes after receiving the interfered versions of each other's messages, first of all tries to locate its own message, which has undergone random delay and attenuation while passing through the channel. To this end, each source node ($S_i$) correlates its own signature and transmitted sequence with the received sequence. The amplitude of the resulting peak, which is due to the node's own message being aligned with its replica in the received sequence, enables us to estimate the channel gain for that particular data packet; whereas the index at which the peak occurs, tells us about the total delay which the data packet has gone through. Finally, knowing the channel function, the source node is able to reconstruct and remove its own message from the received signal and use the remaining parts to decode the other node's message. The BER versus SNR performance of the above system has been simulated in section \ref{sim:dig}.

\subsection{Real Amplitude Scaling}\label{sec:RealNum}
\begin{figure}[tp]
\centering
\includegraphics[totalheight=40mm]{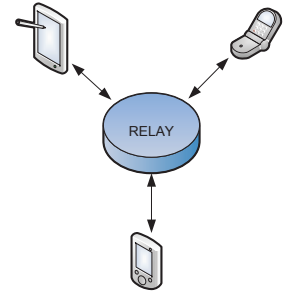}
\caption{The 4-node start topology for wireless relay networks.}
\label{fig:starnet}
\end{figure}

The RAS method enables an arbitrary number of source nodes, arranged in a star topology, to communicate with each other through a relay node, as depicted in Fig. \ref{fig:starnet}. 

As pointed out earlier in section \ref{introduction}, using the RAS approach, while the first signature has a constant preset amplitude ($A \triangleq 1$), the amplitude of the second signature will be scaled according to the transmitted data, $A.r_{i} = r_i$.

\begin{figure}[tp]
\centering
\includegraphics[totalheight=60mm]{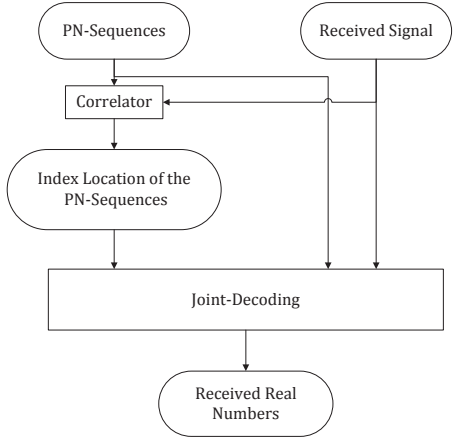}
\caption{The Decoding (Reception) procedure for the RAS System.}
\label{fig:BD}
\end{figure}

The block-diagram in Fig. \ref{fig:BD} depicts how each source node decodes and interprets the interfered signal, forwarded by the relay node.

Suppose that the $j-$th node ($S_j$) is decoding its received signal. First, the ``Correlator" uses the first signature of each of the source nodes,  $S_i, \; i = 1, \ldots, N$, to locate the corresponding messages in the received signal. The ``Joint-Decoding" module then divides the received signal's sequence into several segments, starting from the beginning of the received stream. Knowing where each of the interfering PN sequences is located, we are able to determine which signals have been summed together to produce each segment. This way, we can correspond a linear algebraic equation to each of these segments and the solution of the resulting linear system of equations will provide us with enough information to determine each of the transmitted real numbers. 

Our goal is to make all real numbers, $r_i, i = 1,\ldots,N \; , i \neq j $,  transmitted by the rest of the source nodes, known to the decoding node, $S_j$. In order to express the contribution made by $S_i$'s transmitted sequence in $S_j$'s received sequence, we use two unknowns: $z_{ij}$ and $x_{ij}$. They correspond to the contributions made by $S_i$'s first and second signatures, respectively. Using the terminology introduced in equation \ref{channel} the amplitude by which $S_i$'s first signature has taken part in the generation of the received signal will be equal to $z_{ij} = A. g_{ij} = g_{ij}$. Similarly, for the second signature we have $x_{ij} = r_{i}.g_{ij}$. Consequently $S_j$ will decode the values for the transmitted real numbers as:
\begin{equation}
\hat{r}_{ij} = \frac{x_{ij}}{z_{ij}}, i = 1,\ldots,N \; , i \neq j
\label{final}
\end{equation}

Ideally we desire $\hat{r}_{ij}=r_i$ in \ref{final}. According to what was mentioned, we have two sets of $N-1$ unknowns, namely: 
\begin{equation}
\mathbf{Z} = \left[z_{1j},z_{2j},\ldots ,z_{ij},\ldots ,z_{Nj} \right]^T  \; , i \neq j
\label{Z}
\end{equation}
and
\begin{equation}
\mathbf{X} = \left[x_{1j},x_{2j},\ldots,x_{ij},\ldots,x_{Nj} \right]^T   \; ,   i \neq j
\label{X}
\end{equation}
where N is the number of source nodes. 

In the following paragraphs, we will introduce a system of algebraic equations, of the form:
\begin{equation}
C\times \left(\begin{array}{c}\mathbf{Z}\\\mathbf{X}\end{array}\right) - \mathbf{Y} = 0
\label{sys}
\end{equation}
which we will then solve for the unknown vectors $\mathbf{Z}$, and $\mathbf{X}$.
Matrix $C = (C_{kl})_{M\times(2N-2)}$ has $2 \times (N-1)$ columns equal to the total number of unknowns. The number of rows in $C$ is equal to the total number of equations, $M$, and is determined by the length of the received sequence as well as that of the individual segments that we consider in it. $\mathbf{Y} = (y_k)_{M\times1}$ is a column vector with the same number of rows as $C$. The derivation of $\mathbf{Y}$ along with $C$ will be elaborated upon in the next paragraph. 

Starting from the beginning of the received signal we consider each of its segments, $k = 1, \ldots, M$, separately, and assign an equation of the form described in \ref{sys} to that segment. To this end, we need to determine the right hand side of the equation $y_k$, as well as the coefficients for that equation: $C_{kl}, l = 1,...,2\times (N-1)$. $y_k$ is the sum of all samples in the $k-$th segment of the received sequence of samples. As for the coefficients matrix $C$, its elements are determined based on the position of $S_i$'s signatures in the received signal, with respect to the location of the segment under consideration. Three cases may arise according to which the elements of $C$ may be assigned. Table \ref{C_elements} summarizes these cases, and the corresponding elements of $C$.

\begin{table}[tp]
\renewcommand{\arraystretch}{1.3}
\centering

    \caption{Determining the Elements of Matrix $C$}
    \label{C_elements}

    \begin{small}
    \begin{tabular}{|c|c|c|}
    \hline
    {\bf{Case}} & {$\mathbf{C(k,i)}$} & {$\mathbf{C(k,2i)}$}\\           
    \hline
    \parbox[t]{25mm}{\centering $S_i$'s signatures do not overlap with the $k-$th segment} & 0 & 0\\
    \hline
    \parbox[t]{25mm}{\centering Only $S_i$'s first signature overlaps with the  $k-$th segment} & \parbox[t]{25mm}{\centering Sum of all the samples in the overlapping part of $1^{st}$ signature} & 0\\

    \hline
    \parbox[t]{25mm}{\centering Only $S_i$'s second signature overlaps with the  $k-$th segment} & 0 & \parbox[t]{25mm}{\centering Sum of all the samples in the overlapping part of $2^{nd}$ signature} \\
    \hline
    \end{tabular}
    \end{small} 
\end{table}
 
Finally with all the coefficients available, we solve the resulting overdetermined system of equations in \ref{sys}, for a least squares error solution:
\begin{equation}
\left(\begin{array}{c}\mathbf{Z}\\\mathbf{X}\end{array}\right) = {(C^TC)}^{-1}C^T \times \mathbf{Y}
\label{sysLS}
\end{equation}

It is worth highlighting that the choice for the length of segments poses an interesting trade-off: the longer the length of segments, the less noisy each of the equations; however, the total number of the equations (thus the data at hand) will be smaller. Theoretically though, regardless of the length of the segments, they should all lead to the same functionality, as changing the length of the segments does not have any implications on the amount of the data available. Put in other words, a small number of equations with less noise in each yield the same result as a larger number of them, each with more noise as compared to the previous set. 

We will end the discussion of the RAS decoding scheme by an allusion to the way a source node's knowledge of its own transmitted number, $r_j$, has been exploited in our system. However, to remove the node's message from the received signal as in section \ref{sec:digitEnc}, we need to know how this message has been affected by the channel, a piece of information which is unavailable, until the whole ``joint-decoding" process is over. Therefore, some other solution needs to be sought. Presently, what we do is to add an additional equation of the form:
\begin{equation}
r_j . z_{jj} + x_{jj} = 0
\label{extraeq}
\end{equation}
However, whether or not this is the optimum solution remains to be investigated. Nonetheless, there remain issues that are yet to be addressed regarding the larger networks, amongst which the question of whether the relay nodes should amplify-and-forward or decode-and-forward \cite{nuzdah}. Section \ref{sim:RAS} presents the results of simulations for the above system.

\section{Simulation Results and Discussion}\label{sec:sim}
MATLAB\textsuperscript{\textregistered} has been used to simulate the operation of the proposed systems, using the assumptions set forth for the wireless channel in section \ref{introduction}.

\subsection{Digital Encoding}\label{sim:dig}
\begin{figure}[tp]
     \centering
           \includegraphics[width=90mm,totalheight=50mm]{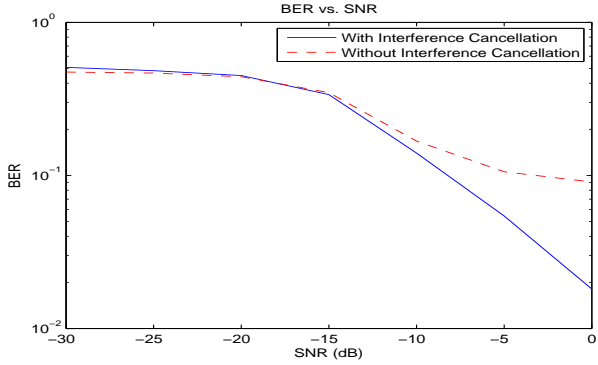}
     \caption{The Performance of the Digital Encoding Scheme with and without Interference Cancellation. The applied signature had 64 chips.}
     \label{fig:digital}
\end{figure}

Fig. \ref{fig:digital} illustrate the BER vs. SNR characteristics of the digital encoding scheme described in section \ref{sec:digitEnc} for the network topology depicted in Fig. \ref{fig:starnet}. In order to evaluate the effectiveness of the proposed scheme, the simulations have been repeated with the ``interference-cancellation" module disabled in the source nodes. The resultant degradation observed in the figure attests to the improvement brought about through interference cancellation.

\subsection{Real Amplitude Scaling}\label{sim:RAS}
\begin{figure}[tp]
     \centering

     \includegraphics[width=90mm,totalheight=50mm]{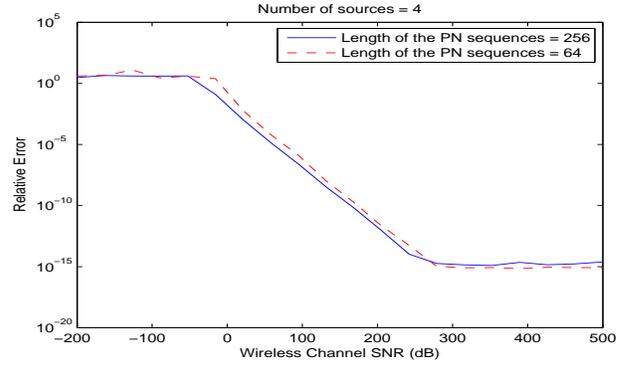}
     \caption{Plotting the Error vs. SNR characteristic over an extended range of SNRs verifies the prediction that the system becomes non-functional for SNRs below a certain limit and the Error is no longer correlated with the Wireless Channel SNR for low SNRs.}
     \label{fig:S3}
\end{figure}

\begin{figure}[tp]
     \centering
           \includegraphics[width=90mm,totalheight=50mm]{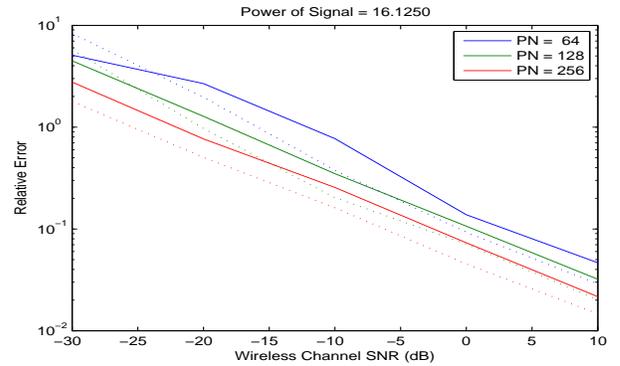}

     \caption{Error vs. SNR  for a star network topology with three (solid) and four (dotted) source nodes.  The Instantaneous Power is set constant at each source node.}
     \label{fig:S2}
\end{figure}

The plots in Fig. \ref{fig:S3} and \ref{fig:S2} are the results of running the RAS system with different number of source nodes and signature lengths, for various values of wireless channel SNR. 
With about 1\% relative error at $SNR=0$dB, for $3$ source nodes and $128$-chip PN-sequences in Fig. \ref{fig:S2}, the system operates seamlessly at moderate and high SNRs. However, when the power of noise is so high that the system is not able to locate the markers in the received data, the value of error is not valid for theoretical analysis. In other words, the system can be expected to exhibit a random behavior for very low SNRs or large number of source nodes. We may refer to the system at the latter condition as non-functional. These operational limits were verified by the results of simulations in Fig. \ref{fig:S3}, where at SNRs bellow $-16$dB, the Error magnitude became independent of SNR and remained almost constant at $2.5$; however, a linearly decreasing pattern was observed in the logarithmic scale for the SNRs above $-15$dB, until it saturated at about $300$dB, because of the computer round-off error. Fig. \ref{fig:S3} further shows that increasing the length of the signatures enhances the system's performance by increasing its resilience to the channel noise, and decreasing the lower limit on SNR. The lower limit on SNR is decreased from $-16$dB in the case of four source nodes and $64$-chip PN-sequences to $-53$dB for the same number of source nodes and $256$-chip PN-sequences.

\subsection{Comparison of the Digital and Analog Network Coding Schemes}\label{sim:comparison}
The performance of the traditional routing, digital network coding and the proposed analog scheme are simulated for a three node chain network and the results are included here. The total length of communication is fixed for all scenarios. Therefore, the Analog Network Coding (ANC), Digital Network Coding(DNC) and traditional routing approach will use the available time in two, three and four time slots respectively. Moreover, the instantaneous power available to each source node is constant and fixed for all nodes. The message transmitted by a source node using the ANC approach is constructed as bellow: 
\begin{equation}
\mbox{ANC Message} = [A \times \mbox{(Marker)}, X\times (\mbox{Signature})]
\label{ANCmessage}
\end{equation}
where Signature and Marker are random binary ($\pm 1$) sequences with equal length. Since the real number, which has normal distribution should be scaled to the value suitable for transmission, $X$ is defined as follows:
\begin{equation}
X =\frac{((R \times 100)+40)}{40}
\label{scaling}
\end{equation}
where $R$ is the normally distributed random real number that is to be communicated.
The power of ANC Sequence can thus be computed as:
\begin{eqnarray}
\mbox{Power ANC} &=& \frac{1}{\mbox{total length}} \times \{A ^2 \times \mbox{length(Marker)}     \nonumber \\ 
								 &+& \varepsilon \left[ X^2 \right] \times \mbox{length(Signature)} \} \nonumber \\
                 &=& \frac{1}{2} \times \{ A^2 + (({2.5})^2 \times \varepsilon \left[ R^2 \right] +1 )  \} \nonumber  \\
                 &=& 0.5 \left( A^2 \right)  +  3.625 
\label{powerANC}
\end{eqnarray}
where $\varepsilon[.]$ denotes the mathematical expectation.

The computation of the instantaneous power for the digital techniques is straightforward and is given by: $A^2$. The digital schemes use markers for frame synchronization to indicate the beginning of packets.

\begin{figure}[tp]
     \centering
     \subfigure[]{               \label{fig:ANC90}\includegraphics[width=90mm,totalheight=50mm]{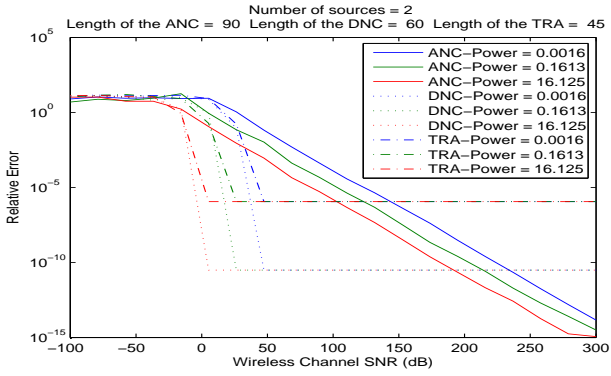}}

     \subfigure[]{
\label{fig:ANC90_lowSNR}\includegraphics[width=90mm,totalheight=50mm]{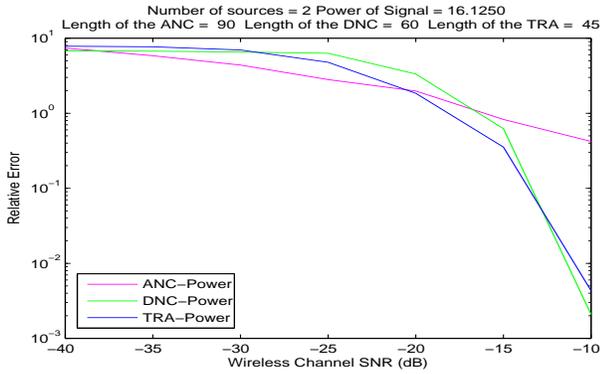}}
     
\caption{Comparing the performance of the three routing schemes. The operation is simulated for a three node chain network, where a relay node connects two source nodes. Available time of communication is set at 180 chips.(a) Extended range (b) Low SNRs}

\label{fig:digitalAnalogCompare}
\end{figure}

Fig. \ref{fig:digitalAnalogCompare} shows the simulations results. It is worth noting that while the operation of the ANC scheme is limited at the higher SNRs by the computer precision ($\sim 10^{-15}$), the error plot for digital schemes will be saturated at the value of error equal to the their quantization error, which decreases with the increasing time of communication and is independent of wireless channel SNR. Finally, Fig. \ref{fig:ANC90_lowSNR} shows the superiority of the analogue approach at low SNRs. 

\section{Conclusion}\label{sec:conc}
Network codes allow intermediate nodes in the network to ``mix" information, typically over finite fields, resulting in improved performance compared to routing. However, sometimes it is more ``natural" to mix over the real field, instead of finite fields. For instance, interference in wireless networks naturally results in real or complex linear combinations of the transmitted signals. Also, for data sources like voice, images, and video, the information consists of sequences of real numbers. We believe that the proposed RAS approach provides a perfect framework for implementation of network codes over real fields. Moreover, the RAS approach offers a straightforward strategy for the treatment of extra information, such as reception of multiple versions of the same message through either multiple transmissions of the same signal or multiple-paths, by considering them as additional equations that can be readily incorporated into the existing system of linear equations. This paper's primary focus was the structure and design of the proposed algorithm, postponing a complete theoretical analysis of its performance, structure, computational complexity and so on until a future study.

\section*{Acknowledgment}
The authors would like to thank L. Ghabeli for her help with this work.
\nocite{*}
\bibliographystyle{IEEEtran}
\bibliography{ICC2011}

\end{document}